\documentclass[11pt]{article}
\usepackage{geometry}
\usepackage{amsmath}
\usepackage{amssymb}
\usepackage{amsthm}
\usepackage{algorithm,algpseudocode}
\usepackage{color}
\usepackage{booktabs} 
\usepackage{graphicx}
\usepackage{url}
\usepackage{setspace}

\theoremstyle{thmstyleone}%
\newtheorem{theorem}{Theorem}

\theoremstyle{thmstyletwo}%
\newtheorem{example}{Example}%

\theoremstyle{thmstylethree}%
\newtheorem{definition}{Definition}%

\newtheorem{assumption}{Assumption}
\newtheorem{problem}{Problem}
\newtheorem{corollary}{Corollary}

\newcommand{\prob}{\operatorname{prob}}
\newcommand{\dv}{\operatorname{DV}}

\newcommand{\pa}{\operatorname{PA}}
\newcommand{\pc}{\operatorname{PC}}
\newcommand{\ate}{\operatorname{ATE}}
\newcommand{\cate}{\operatorname{CATE}}

\renewcommand{\emph}{\textit}

\newcommand{\Cross}{\mathbin{\tikz [x=1.4ex,y=1.4ex,line width=.2ex] \draw (0,0) -- (1,1) (0,1) -- (1,0);}}%

\makeatletter
\newcommand*{\indep}{%
	\mathbin{%
		\mathpalette{\@indep}{}%
	}%
}

\newcommand*{\nindep}{%
	\mathbin{
		\mathpalette{\@indep}{/}%
	}%
}

\newcommand*{\@indep}[2]{%
	\sbox0{$#1\perp\m@th$}
	\sbox2{$#1=$}
	\sbox4{$#1\vcenter{}$}
	\rlap{\copy0}
	\dimen@=\dimexpr\ht2-\ht4-.2pt\relax
	\kern\dimen@
	\ifx\\#2\\%
	\else
	\hbox to \wd2{\hss$#1#2\m@th$\hss}%
	\kern-\wd2 %
	\fi
	\kern\dimen@
	\copy0 
}
\makeatother

\begin{document}
		
		\title{Causal heterogeneity discovery by bottom-up pattern search for personalised decision making}
		
		\author{Jiuyong Li$^1$ \and Lin Liu$^1$ \and Shisheng Zhang$^1$ \and Saisai Ma$^2$ \and Thuc Duy Le$^1$ \and Jixue Liu$^1$ \\
		$^1$STEM, University of South Australia, Adelaide, Australia \\
	$^2$Consilium Technology, Australia}

		



		


		\maketitle

\abstract{
	In personalised decision making, evidence is required to determine whether an action (treatment) is suitable for an individual. Such evidence can be obtained by modelling treatment effect heterogeneity in subgroups. The existing interpretable modelling methods take a top-down approach to search for subgroups with heterogeneous treatment effects and they may miss the most specific and relevant context for an individual. In this paper, we design a \emph{Treatment effect pattern (TEP)} to represent treatment effect heterogeneity in data. 
	To achieve an interpretable presentation of TEPs, we use a local causal structure around the outcome to explicitly show how those important variables are  used in modelling. We also derive a formula for unbiasedly estimating the \emph{Conditional Average Causal Effect (CATE)} using the local structure in our problem setting.  
	In the discovery process, we aim at minimising heterogeneity within each subgroup represented by a pattern. We propose a bottom-up search algorithm to discover the most specific patterns fitting individual circumstances the best for personalised decision making. Experiments show that the proposed method models treatment effect heterogeneity better than three other existing tree based methods in synthetic and real world data sets.}


\section{Introduction}

We  study the problem of identifying the \emph{Treatment effect patterns (TEPs)} which specify subgroups where a treatment has a significant effect on the outcome. 
For example, chemotherapy is a common cancer treatment, but it is not suitable for all patients. Finding TEPs indicating the types of patients who are benefited (or least benefited) from chemotherapy treatment will be helpful for personalised medicine. For personalised marketing, it will be helpful to identify TEPs indicating the subgroups of customers who will buy a certain product due to a promotion (treatment).    

TEPs are different from the discriminative patterns in data mining literature, e.g. emerging patterns~\cite{emergingPatterns}, contrast sets~\cite{ContrastPatterns} and subgroups~\cite{LocalModelSubgroup,Li-nonredundantSubgroup}. Discriminative patterns specify subgroups where the distribution of the outcome is significantly different from that outside the subgroups, and they are used for classification. For example, the discriminative pattern: \{family background = business\} defines a subgroup where the probability of high income (an outcome for illustration only) is larger than that outside the group. The pattern can be used to predict if a person has a high income or not. 

TEPs are not aimed at predicting an outcome, but are aimed at determining whether to take a treatment (or an action) in decision making. 
%
TEPs take a fixed pair of the treatment and the outcome variables, and represent subgroups where a change in the treatment variable makes a significant change in the outcome. For example, let college education be the treatment and salary be the outcome. The discriminative pattern \{family background = business\} is not a TEP, as for this subgroup college education would not change their income much (this subgroup of people are likely to have high income anyway based on their family background). A TEP would be \{family background = illiterate\}. For this subgroup of people, education can make a big impact on their future careers. 
For example, without a college education, this subgroup of people may nearly all receive a very low salary. After the education, 30\% of individuals in this subgroup receive a salary higher than the median salary in the population. 30\% is lower than 50\%, the expected percentage of the population having a salary above the median. But for this subgroup, 30\% is a big improvement. So this TEP provides strong evidence for personalised decision making on  going to college or not.   


A summary of the differences between TEPs and discriminative patterns is shown in Table~\ref{tab_contrast}. 

\begin{table*}[t]
\footnotesize
	\caption{Differences between TEPs \& discriminative patterns}
	\begin{center}
		\begin{tabular}{| l | p{5.8 cm} | p{7.4 cm} |}
			\hline
			& Discriminative patterns  		&  TEPs~(w.r.t pair $(W, Y)$)  \\
			& $\{W=1, X_1=1\} \to Y=1$ 		&  $\{(X_1=1, X_2=*), X_3=*\}$ \\
			\hline
			Nature 	& Association between $\{W=1, X_1=1\}$ \& $Y=1$	& Causation: changing $W$ leads to a change in $Y$ when $X_1=1$. \\
			\hline
			Semantics & Influence of other variables on the pattern is not indicated. & $(X_1, X_2, X_3)$ are direct causes of $Y$ and $(X_1, X_2)$ confound $(W,Y)$. \\
			\hline
			Usability & Classification & Personalised decisions \\
			\hline
		\end{tabular}
	\end{center}
	\label{tab_contrast}
\end{table*}%

TEPs are different from high utility patterns~\cite{UtilityPatterns21,MiningUtilityPatternHillCliming} studied in recent years. Utility patterns are frequent itemsets (attribute value sets) with the minimum utility based on an internal or external utility measure, whereas TEPs present conditions for a causal relationship between the treatment and the outcome being strong or weak. Utility patterns can be extended to utility rules, but the utility rules capture associations, not causations. 

TEPs are designed for personalised decision making. For example, a TEP in an e-commerce application, (new customer = true, multiple channels = true) with the treatment effect of 0.2 (treatment: sending promotional emails; outcome: visiting the online catalogue within one week) provides the company evidence for targeting this group of customers for email promotion since the email promotion causes the online catalogue visit. In a medical application, a TEP (MFAP3L = low, AGR2 = low, ABCC2 = low), where  MFAP3L, AGR2, and ABCC2 are genes and the low indicates a gene expression level, with the treatment effect of -0.16 (treatment: chemotherapy; outcome: the survival rate) will discourage a doctor to recommend a patient matching the pattern for a chemotherapy treatment since the treatment does not lead to a positive outcome for this group of patients.

Our work is closely related to treatment effect heterogeneity modelling~\cite{AtheyImbens2016_PNAS,kunzel2019metalearners,CHM-Evaluation,Shalit2016,yoon2018ganite,CEVAE2017,Yao2018_Twin}, an active research area in causal inference. We refer readers to the Related Work section for more discussions. Here we focus on tree based modelling methods since we are interested in interpretable modelling considering that interpretation is also crucial in decision making.  

Treatment effect heterogeneity modelling is mainly about Conditional Average Treatment Effect (CATE) estimation which needs the causal graph underlying the data. Most existing works do not explicitly use a causal graph. For example, many works assume a given covariate set $\mathbf{X}$, such as in~\cite{AtheyImbens2016_PNAS,su_subgroup_2009}. Firstly, this covariate set is unknown to users. Secondly, even if a covariate set can be found by another algorithm (see discussions in the Related Work section.), 
the contributions of different variables in a covariate set to CATE estimation are different. For example, confounders which affect both $W$ and $Y$ need to be adjusted in treatment effect estimation, whereas effect modifiers which affect $Y$ but do not affect $W$~\cite{vanderweele2007four} do not need to be adjusted in treatment effect estimation but to be conditioned on. Such differentiation is only possible when the causal graph (or local causal structure) is presented. 

In our pattern representation, we explicitly represent patterns in a local causal graphic structure and this makes causal semantic clearer. We have also proposed to use a local structure search (instead of a global structure search which can be inefficient) to find the two sets of variables in our problem setting: one set to represent confounders of the treatment and outcome, and the other set to denote effect modifiers of the outcome since two sets play different roles in causal effect estimation.  Another advantage of having an explicit presentation of the local causal structure is that users can use their domain knowledge to validate the discovered TEPs since a valid causal graph is supposed to be consistent with the domain knowledge. Such pattern presentation improves the interpretability and usability of a causal effect heterogeneity model greatly.

Tree based methods have been adapted for interpretable causal heterogeneity modelling~\cite{su_subgroup_2009,AtheyImbens2016_PNAS}. These methods employ a top-down approach to recursively split a (sub)population into subgroups with different treatment effects. Their subgroup search is restricted by the choice of root node since all paths include the root, and this limits their capability for capturing significant heterogeneous subgroups.  
%

In this paper, we employ a bottom-up search approach for identifying TEPs (subgroups), starting from the most specific patterns described by the set of all direct causes of the outcome. The patterns with small numbers of records are merged to be statistically significant. The merging process is implemented by  generalisation which aims at minimising heterogeneity within a subgroup of a generalised pattern while maximising the specificity of patterns. When using the discovered TEPs, the most specific pattern matching a person's situation is used for personalised decision making.

The contributions of our work are summarised in the following:
\begin{enumerate}
	\item We design a new representation for causal effect heterogeneity modelling, TEPs, which explicitly represent the local causal structure for interpretable modelling and evaluation.
	\item We derive solutions to use the local causal structure for unbiased CATE estimation in our problem setting. 
	\item We develop a bottom up generalisation algorithm to discover TEPs by considering within pattern homogeneity and pattern specificity. The bottom up approach ensures that the most specific pattern is used for predicting CATE for an individual. 
\end{enumerate}

\section{Problem definition}\label{section_problem_definition}

Let $D$ be a data set containing $n$ records of the triple $(W, Y, \mathbf{X})$, where $W$ is the treatment variable, $Y$ the outcome variable, and $\mathbf{X}$ the set of pretreatment variables representing background conditions and/or characteristics of an individual, denoted by a record in $D$.  Pre-treatment variables are not influenced by $W$ or $Y$ but may influence $W$ or $Y$. We assume that $W$ has an effect on $Y$. $W$ takes two values, 1 and 0, standing for treatment and control respectively. and $Y$ is a binary variable. 

We are interested in answering the question: \emph{``For a subgroup of individuals, will they benefit from receiving the treatment ($W=1$)?''}




What we need is to estimate CATE, i.e. the change of $Y$ as a result of changing or intervening on $W$ under condition $\mathbf{X=x}$.  To make the objective formally, we use Pearl's \emph{do} operator~\cite{Pearl2009_Book}, a notation commonly used in causal inference literature, to represent an intervention. The \emph{do} operation mimics setting a variable to a certain value (not just passively observing a value) in a real world experiment. The probability given a \emph{do} operation, e.g. $\prob(y \mid do (W = 1))$, indicates the probability of $Y=1$ when $W$ is set to 1, and is different from $\prob(y \mid W = 1)$, the probability of $Y=1$ when observing $W=1$. 

Let $\mathbf{P=p}$ (or simply $\mathbf{p}$) where $\mathbf{P \subseteq X}$ be a pattern which represents a subgroup in the population. For example, (male, professional) is a pattern representing a type of employees. CATE associated with pattern $\mathbf{p}$ is defined as the following.
\begin{eqnarray}
\centering
\label{CATE}
& CATE_{\mathbf{p}}(W,Y) = & \prob(y \mid do (W = 1), \mathbf{p}) -  \nonumber \\
& & \prob(y \mid do (W = 0), \mathbf{p})
\end{eqnarray}

When $\mathbf{P}$ is an empty set, $\cate_{\emptyset}(W,Y) $ is the Average Treatment Effect (ATE) in the population, specifically. 
\begin{eqnarray}
\label{ATE}
&  ATE(W,Y) =&  \prob(y \mid do (W = 1)) - \nonumber \\
 & & \prob(y \mid do (W = 0)) 
\end{eqnarray}

To make CATE estimation close to the individual level, we need $\mathbf{p}$ to be specific. However, the estimated CATE may not be reliable when there are a small number of samples in the subgroup of $\mathbf{p}$. Given a data set, a pattern cannot be too specific since its CATE estimation has to be reliable. In contrast, a general pattern may contain heterogeneous treatment effects within its subgroup. Putting both considerations together, we have the following problems to be tackled in this paper.
\begin{definition}[Problem definition]
	Given a data set $D$ of $(W, Y, \mathbf{X})$ and $\mathbf{X}$ is a pretreatment set of $(W,Y)$, we aim to design and find a set of patterns for personalised decision making. A pattern should be as specific as possible while its subgroup should be large enough for reliable CATE estimation. The CATEs of sub subgroups in the subgroup should have as a small difference as possible.
\end{definition}

Equation~\ref{CATE} is conceptual and the CATE of a pattern cannot be estimated directly from data yet. Our next step is to develop an analytic expression to estimate CATE for a pattern from data. Firstly, we will introduce the background of causal graphs and the calculus of intervention.

\section{Causal DAG and $do$ calculus} 
\label{sec_causalDAG}

A DAG (Directed Acyclic Graph), denoted as $\mathcal{G}=(\mathbf{V},\mathbf{E})$, is a directed graph where $\mathbf{V}$ contains a set of nodes, $\mathbf{E}$ contains a set of directed edges, and no node has a sequence of directed edges pointing back to itself. If there exists an edge $V_1 \rightarrow V_2$ in $\mathcal{G}$, $V_1$ is a parent node of $V_2$ and $V_2$ is a child node of $V_1$.  We use $\pa(V_2)$ to denote the set of all parents of $V_2$. 
A path is a sequence of nodes linked by edges regardless of their directions. A directed path is a path on which all the edges follow the same direction. Node $V_1$ is an ancestor of node $V_2$ if there is a directed path from $V_1$ to $V_2$, and equivalently $V_2$ is a descendant of $V_1$. $V_2$ is a collider if $V_1 \to V_2 \leftarrow V_3$. 
\begin{definition}[Markov condition~\cite{Pearl2009_Book}]
	Let $\mathcal{G}=(\mathbf{V},\mathbf{E})$ be a DAG and $P(\mathbf{V})$ be the probability distribution over $\mathbf{V}$.
	$P(\mathbf{V})$ and $\mathcal{G}$ satisfy the Markov condition if, $\forall V\in\mathbf{V}$, $V$ is conditionally independent of all of its non-descendants given $\pa(V)$.   
\end{definition}
When the Markov condition holds, the joint distribution of $\mathbf{V}$ is factorised as $\prob (\mathbf{V}) =
\prod_{V_i \in \mathbf{V}} \prob(V_i \mid \pa (V_i))$. 
\begin{definition}[Faithfulness\cite{Spirtes2000_Book}]
	If all the conditional independence relationships in $P(\mathbf{V})$ are entailed by the Markov condition applied to DAG $\mathcal{G}=(\mathbf{V},\mathbf{E})$, and vice versa, $P(\mathbf{V})$ and $\mathcal{G}$ are faithful to each other.     
\end{definition}
The faithfulness assumption is to ensure that the DAG $\mathcal{G}=(\mathbf{V}, \mathbf{E})$ represents all the conditional independence relationships in the joint distribution $P(\mathbf{V})$ and vice versa. 

The following causal sufficiency assumption is needed when estimating treatment effects in data in addition to the Markov condition and faithfulness assumption. 
\begin{definition}[Causal sufficiency~\cite{Spirtes2000_Book}]
	For every pair of variables observed in a data set, all their common causes are also observed in the data set.
\end{definition}
%
%
Given the three assumptions, a DAG learned from data 
is a causal DAG, and parents are interpreted as the direct causes of their children.

$d$-Separation as defined below is an important concept to read the conditional independences/dependencies among nodes from a causal DAG. 
\begin{definition} [$d$-Separation~\cite{Pearl2009_Book}]
	A path $p$ in a DAG is $d$-separated by a set of nodes $\mathbf{Z}$ if and only if 
	(1) $\mathbf{S}$ contains the middle node, $V_k$ of a chain $V_i \to V_k \to V_j$, $V_i \leftarrow V_k \leftarrow V_j$, or a fork  $V_i \leftarrow V_k \to V_j$ in $p$; and 
	(2) when $p$ contains a collider $V_k$, i.e. $V_i \to V_k  \leftarrow V_j$, none of $V_k$ and its descendants is in $\mathbf{S}$.
\end{definition}
When all paths between $V_1$ and $V_2$ are $d$-separated by $\mathbf{S}$ in a DAG, we have $(V_1 \indep V_2 \mid \mathbf{S})$. We call $\mathbf{S}$ \emph{blocks} a set of paths if it $d$-separates all the paths simultaneously.  



Now we use DAG for causal effect estimation. 
\begin{definition}[The backdoor criterion~\cite{Pearl2009_Book}]
	Given a causal DAG $\mathcal{G}=(\mathbf{V}, \mathbf{E})$, for an ordered pair of variables $(W, Y)$ in $\mathbf{V}$, a set of variables $\mathbf{Z}\subseteq \mathbf{V}\setminus\{W, Y\}$ is said to satisfy the backdoor criterion if (1) $\mathbf{Z}$ does not contain a descendant node of $W$, and (2)  $\mathbf{Z}$ d-separates every path between $W$ and $Y$, containing an arrow into $W$. 
\end{definition}
%

Once a set $\mathbf{Z}$ satisfying the backdoor  criterion with respect to the variable pair $(W, Y)$ is identified. $\prob (y \mid do (W=w), \mathbf{Z})$ is reduced to $\prob (y \mid W=w, \mathbf{Z})$ where $w \in \{0, 1\}$. This means that the causal effect defined by $do()$ operations can be estimated in data. The set $\mathbf{Z}$ is called an \emph{adjustment} (or deconfounding) set relevant to $(W, Y)$.

$do$-calculus rules~\cite{Pearl2009_Book} are more general criteria for reducing a $do()$ operation to a normal statistical expression, and are used in our derivations of CATEs for patterns.  The $do()$ operation on a variable, e.g. $do (X=x)$ in DAG $\mathcal{G}$ can be represented by removing all incoming edges of $X$ from $\mathcal{G}$. Let $V_1$ and $V_2$ be two variables in $\mathcal{G}$. $\mathcal{G}_{\overline{V_1}}$ represents the mutilated graph of $\mathcal{G}$ by removing all incoming edges of $V_1$, $\mathcal{G}_{\underline{V_2}}$ the mutilated graph of $\mathcal{G}$ by removing all outgoing edges of $V_2$, $\mathcal{G}_{\overline{V_1}, \overline{V_2}}$ the mutilated graph of $\mathcal{G}$ by removing all incoming edges of $V_1$ and $V_2$, and $\mathcal{G}_{\overline{V_1}\underline{V2}}$ the mutilated graph of $\mathcal{G}$ by removing all incoming edges of $V_1$ and all outgoing edges of $V_2$. When $\mathbf{V_1}$ or $\mathbf{V_2}$ represents a variable set, the edge removal is then for each variable in the set. The rules of $do$-calculus are given as Theorem 3.4.1 in~\cite{Pearl2009_Book}.

\section{Bottom up discovery of TEPs}

\subsection{CATE estimation in the local causal structure}

\begin{figure}[tb]
	\begin{center}
		\includegraphics[width=0.27\textwidth]{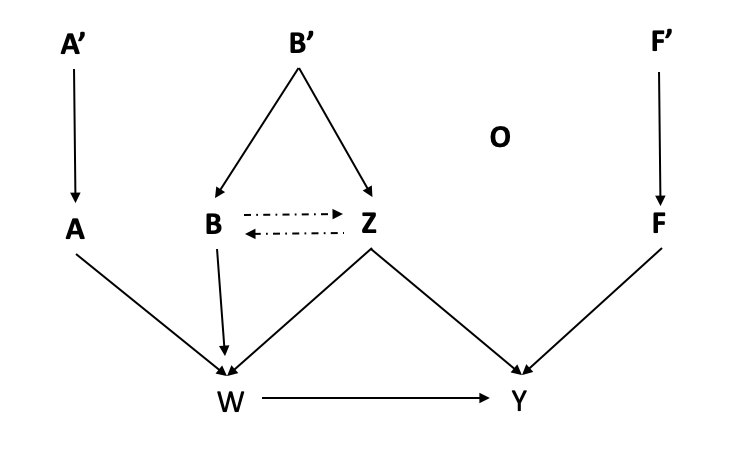}
		\caption{An illustrative causal DAG in the problem setting. Two dash edges mean alternative paths, either into $\mathbf{Z}$ or $\mathbf{B}$ (not both). $\mathbf{A, A', B, B', Z, F, F'}$ and $\mathbf{O}$ are sets and there are edges between variables in a set. Edges between two sets can be multiple although only one edge is shown.  }
		\label{fig_CausalStructure}
	\end{center}
\end{figure}

An exemplar sketch of causal DAG in the problem setting is shown in Figure~\ref{fig_CausalStructure}. $\mathbf{A, A', F, F'}$ are parents and ancestors of $W$ and $Y$ respectively.  $\mathbf{B, B', Z}$ are parents and/or ancestors of both $W$ and $Y$. $\mathbf{Z}$ is an adjustment set of $(W, Y)$ (to be discussed later in this section). $\mathbf{O}$ contains irrelevant variables which are independent of both $W$ and $Y$.

In Section~\ref{section_problem_definition}, pattern $\mathbf{p}$ is defined as a value assignment of set $\mathbf{P \subseteq X }$. Based on the causal graph and do-calculus, we have the following refinement.
%
%
%
\begin{theorem}
	\label{theorem_parent}
	Given a variable pair $(W, Y)$ and a set of pretreatment variables $\mathbf{X}$. $W$ has a treatment effect on $Y$.  Patterns defined in $\pa'(Y)$ where $\pa'(Y) = \pa(Y) \backslash W$ capture all treatment effect heterogeneities of $W$ on $Y$ defined by $\mathbf{X}$ and a superset of $\pa'(Y)$.    
\end{theorem}
\begin{proof}
	Let us consider a pattern $\mathbf{X=x}$. Based on the definition, $\cate_{\mathbf{x}}(W,Y) =  \prob(y \mid do(W = 1), \mathbf{x}) - \prob(y \mid do(W = 0), \mathbf{x})$.
	
	Let $w$ be a value of treatment $W$. Since the two terms of $\cate_{\mathbf{x}}(W,Y)$ are the same except for the values of $W$. We show how the expression with $do(w)$ is simplified.  
	
	Let $X = \pa'(Y) \cup \mathbf{N}$ where  
	$\pa'(Y) \cap \mathbf{N} = \emptyset$. $\mathbf{N}$ represents the set of all non-parent nodes of $Y$. Let $N_1$ be a variable in  $\mathbf{N}$, and $\mathbf{N'} = \mathbf{N} \backslash N_1$.  We have the following reduction. 
	\begin{eqnarray*}
	&& \prob(y \mid do(w), \mathbf{x})  \\ &=&   \prob(y \mid do(w), \pa'(Y) 
	= \mathbf{p}, \mathbf{N' = n'}, N_1 = n_1) \\ &=&  \prob(y \mid do(w), \pa'(Y)=\mathbf{p}, \mathbf{N'=n'})\\ &&\textnormal {(do calculus rule 1 in Theorem 3.4.1 \cite{Pearl2009_Book})} 
	\end{eqnarray*}
	In the above reduction, the following rationale is used. Firstly, if there are one or more paths linking $N_1$ to $Y$ in the mutilated graph $\mathcal{G}_{\overline{W}}$ where all the incoming edges of $W$ are removed. $N_1$ is $d$-separated from $Y$ by $\pa'(Y)$ in $\mathcal{G}_{\overline{W}}$. Hence, $N_1 \indep Y \mid \pa'(Y)$ (equivalently $N_1 \indep Y \mid \pa'(Y), \mathbf{N'}$ since there are no colliders at $\mathbf{N'}$ in $\mathcal{G}_{\overline{W}}$. Therefore, $N_1$ is removed from the equation based on $do$ calculus rule 1 in Theorem 3.4.1~\cite{Pearl2009_Book}. Secondly, if there is not a path linking $N_1$ to $Y$ in $\mathcal{G}_{\overline{W}}$, $N_1 \indep Y \mid \pa'(Y), \mathbf{N'}$ trivially holds and hence the $do$ calculus rule 1 is applied. 
	
	By repeatedly using  Rule 1 in Theorem~3.4.1~\cite{Pearl2009_Book}, all variables in $\mathbf{N'}$ are removed from the equation one by one. We obtain the following equation. 
	
	$\prob(y \mid do(w), \mathbf{x})  =   \prob(y \mid do(w), \pa'(Y)=\mathbf{p})$. 	
	
	So $\cate_{\mathbf{x}}$ is determined by a pattern defined by the parents of $Y$ excluding $W$.
	
	Following the above procedure, any pattern defined by a superset of $\pa'(Y)$ can be reduced to a pattern in $\pa'(Y)$ with the same CATE. 
	
	Hence patterns defined in $\pa'(Y)$ capture all treatment effect heterogeneities defined by $\mathbf{X}$ and a superset of $\pa'(Y)$. 
\end{proof}
Theorem~\ref{theorem_parent} reduces the complexity for finding patterns significantly. This is different from feature selection since $\mathbf{A, A', B, B', Z', F'}$ are all associated with $Y$. The strength of association between two adjacent variables may not be stronger than that between two unadjacent variables. For example, the association between $\mathbf{A}$ (or $\mathbf{A'}$) and $Y$ could be stronger than the association between $\mathbf{Z}$ (or $\mathbf{F}$) and $Y$. So feature selection cannot find the parents of $Y$. 

The parents of $Y$ can be found in a causal graph. In some real world applications, parents of $Y$ are known by domain experts since they are direct causes of $Y$. The parents of $Y$ can be learned in data in our problem setting and we will discuss learning parents in data in Section~\ref{sec_alg}.

$\pa'(Y)$ contains confounders and the parents of $Y$ only. Confounders are variables that affect both (the selection of) treatment $W$ and effect $Y$, and hence need to be adjusted in treatment effect estimation. In graphical terms, Confounders have paths into both $W$ and $Y$ in our problem setting. Let set $\mathbf{Z}$ be parents of $Y$ and parents (or ancestors) of $W$. $\mathbf{F} = \pa'(Y) \backslash \mathbf{Z}$ is the set of parents of $Y$ only, and they do not have paths into $W$. In our problem setting, $\mathbf{F} \indep W$ since variables in $\mathbf{F}$ are not parents or ancestors of $W$. We separate $\mathbf{Z}$ from other variables because of the following property.
\begin{corollary}
	\label{theorem_AjustmentSet}
	Set $\mathbf{Z}$ is a minimal adjustment set for pair ($W, Y$) and the average treatment effect of $W$ on $Y$ is 
	$
	ATE(W,Y) =  \sum_{z}(\prob(y \mid W = 1, \mathbf{z}) - \prob(y \mid W = 0, \mathbf{z})) \prob(\mathbf{z})
	$
\end{corollary}	
\begin{proof}
	Set $\mathbf{Z}$ blocks all the backdoor paths of $(W, Y)$ since $\mathbf{F} = \pa'(Y) \backslash \mathbf{Z}$ do not have backdoor paths into $W$. According to Theorem~\ref{theorem_AjustmentSet}, set $\mathbf{Z}$ is an adjustment set and the $\ate(W, Y)$ is calculated by the summation shown. A subset of $\mathbf{Z}$ leaves a backdoor path unblocked, and does not satisfy Theorem~\ref{theorem_AjustmentSet}. Hence, set $\mathbf{Z}$ is minimal. 
\end{proof}
The parents of $Y$ only (which are $d$-separated from $W$ by an empty set) are effect modifiers, e.g. $\mathbf{F}$. The average treatment effects between $(W,Y)$ conditioned on different values of $\mathbf{F}$ are different~\cite{vanderweele2007four}.

\subsection{The minimal TEP set}  

Now we can define treatment effect patterns to represent the causal heterogeneity in data.

\begin{definition}[Treatment effect patterns (TEPs)]
	Given a variable pair $(W, Y)$ and a set of pretreatment variables $\mathbf{X}$. Let $\mathbf{P} = \pa'(Y) \subseteq \mathbf{X}$. A TEP is a value set $\mathbf{P=p}$ representing a subgroup of population and its associated treatment effect is $\cate(W, Y)_\mathbf{p}$. To represent the local causal structure around $Y$, a TEP is represented as $\mathbf{\{(Z=z), F=f\}}$ where $\mathbf{Z \cup F = P}$, $\mathbf{Z \cap F = \emptyset}$, $\mathbf{Z}$ denotes a set of confounders and $\mathbf{F}$ stand for a set of effect modifiers.    
\end{definition}	 

Let us use $\pa'(Y)=\{X_1, X_2, X_3\}$, $\mathbf{Z} =\{X_1, X_2\}$ and $\mathbf{F}=\{X_3\}$ as an example. $\mathbf{p_1} = \{(X_1=1, X_2=0), X_3=1\}$ is a TEP.

\begin{definition}[Specific and general TEPs]
	A TEP $\mathbf{p}$ is one of the most specific patterns if all its values are specified. A general pattern contains one or more unspecific values `$*$', and represents the union of subgroups of two or more most specific TEPs. When we consider the relationship between two TEPs, we drop unspecified values. If $\mathbf{p_2} \subset \mathbf{p_1}$, TEP $\mathbf{p_2}$ is more general than TEP $\mathbf{p_1}$ or TEP $\mathbf{p_1}$ is more specific than TEP $\mathbf{p_2}$. 
\end{definition}	 

For example, consider $\mathbf{p_1} = \{(X_1=1, X_2=0), X_3=1\}$ and pattern $\mathbf{p_2} = \{(X_1=1, X_2=*), X_3=1\}$. pattern $\mathbf{p_2}$ is more general than pattern $\mathbf{p_1}$ or pattern $\mathbf{p_1}$ is more specific than pattern $\mathbf{p_2}$. 

Note that $X=*$ in a TEP does not mean simply dropping variable $X$ as in the traditional emerging patterns~\cite{emergingPatterns}, contrast sets~\cite{ContrastPatterns} and subgroups~\cite{LocalModelSubgroup,Li-nonredundantSubgroup}  since an unspecified value of a variable in $\mathbf{Z}$ affects the CATE estimation as discussed below. 

Now we  derive $\cate(W, Y)$ when there are unspecified values, i.e. `$*$'s. Let $\mathbf{Z = Z_1 \cup Z_2}$ and $\mathbf{F = F_1 \cup F_2}$ where $\mathbf{Z_1}$ and $\mathbf{F_1}$ contain specified values and $\mathbf{Z_2}$ and $\mathbf{F_2}$ contain unspecified values.

\begin{figure}[t]
	\begin{center}
		\includegraphics[width=0.48\textwidth]{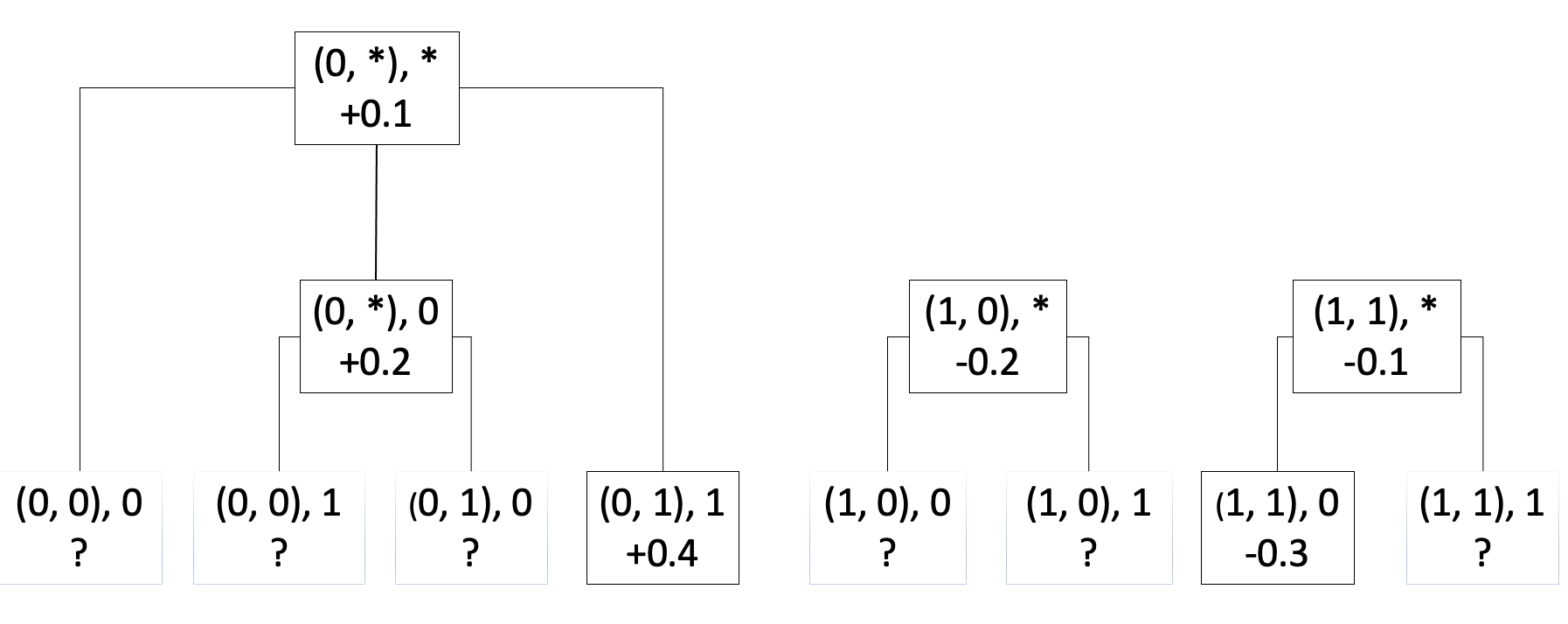}
		\caption{An illustration of the minimal significant TEP set whose members are in the box with the solid line. TEPs with `?' treatment effects are insignificant. This example also shows how TEPs are discovered via pattern generalisation. }
		\label{fig_exampl_TEP_Set}
	\end{center}
\end{figure}

\begin{theorem}
	\label{theorem_CATE_Pattern}
	In the problem setting, $\cate_{\mathbf{p}}(W,Y) = \sum_{z_2}(\prob(y \mid W = 1, \mathbf{z_1, z_2, f_1})$ \\$- \prob(y \mid W = 0, \mathbf{z_1, z_2, f_1})) \prob(\mathbf{z_2 \mid z_1, f_1})$
\end{theorem}

\begin{proof}
	Based on the definition, $\cate_{\mathbf{p}}(W,Y) =  \prob(y \mid do(W = 1), \mathbf{z_1, f_1}) - \prob(y \mid do(W = 0), \mathbf{z_1, f_1})$ since $\mathbf{P = \{ Z_1, F_1\}}$.
	
	Let $w$ be a value of treatment $W$. Since the two terms of $\cate_{\mathbf{p}}(W,Y)$ are the same except for the values of $W$. We will show how $do(W=w)$ (shorted as $do(w)$) is reduced a $do$ free expression. 
	\begin{eqnarray*}
	&&  \prob(y \mid do(w), \mathbf{z_1, f_1}) = \sum_{\mathbf{z_2}} \prob(y, \mathbf{z_2} \mid do(w), \mathbf{z_1, f_1})  \\
	& =&   \sum_{\mathbf{z_2}} (\prob(y \mid do(w), \mathbf{z_1, z_2, f_1})\prob(\mathbf{z_2} \mid do(w), \mathbf{z_1, f_1})) \\
	 & = &  \textnormal{(Rule~2)}  \sum_{\mathbf{z_2}} (\prob(y \mid w, \mathbf{z_1, z_2, f_1})\prob(\mathbf{z_2} \mid do(w), \mathbf{z_1, f_1}))  \\
	& = &  \textnormal{(Rule~3)} \sum_{\mathbf{z_2}} (\prob(y \mid w, \mathbf{z_1, z_2, f_1})\prob(\mathbf{z_2} \mid \mathbf{z_1, f_1})) 
	\end{eqnarray*}
	In the second last step of reduction, $do$ calculus rule 2 in Theorem 3.4.1~\cite{Pearl2009_Book} has been used. In the mutilated graph $\mathcal{G}_{\underline{W}}$ where edge $W \to Y$ is removed, $W$ is $d$-separated from Y by $\mathbf{Z_1, Z_2}$. There are no colliders at $\mathbf{F_1}$. Hence, $W \indep Y \mid \mathbf{Z_1, Z_2, F_1}$ in $\mathcal{G}_{\underline{W}}$ and ``do" is removed from $do(W)$.
	
	In the last step of reduction, $do$ calculus rule 3 in Theorem 3.4.1~\cite{Pearl2009_Book} has been used. In the mutilated graph $\mathcal{G}_{\overline{W}}$ where edges into $W$ have been removed, $W$ is $d$-separated from $\mathbf{Z_2}$ by the empty set. $W$ is $d$-separated from $\mathbf{Z_2}$ by set $\{\mathbf{Z_1, F_1}\}$ since there are no colliders at $\mathbf{Z_1, F_1}$. Hence, $W \indep \mathbf{Z_2} \mid \mathbf{Z_1,  F_1}$ in $\mathcal{G}_{\overline{W}}$ and $do(\mathbf{Z_2})$ is removed from the equation.

	Therefore, the Theorem is proved. 
\end{proof}

The CATE of the most general pattern, such as, $\mathbf{p_3} = \{(X_1=*, X_2=*), X_3=*\}$, is the $\ate(W, Y)$ in the population.



We are interested in significant patterns with reliable statistics. 

\begin{definition} [Significant patterns]
	Pattern $\mathbf{p}$ is significant if the difference $\Delta= \lvert \prob(y \mid W=1, \pa'(Y)=\mathbf{p}) - \prob(y \mid W=0, \pa'(Y)=\mathbf{p}) \rvert $ is  greater than 0 statistically. 
\end{definition}

We use a critical ratio statistic as in~\cite{Fleiss-Rates} to test the significance of difference $\Delta$. Based on the values of $W$ and $Y$, we obtain the following cross table where $n_{*j}=n_{1j}+n_{0j}$, $n_{i*}=n_{i1}+n_{i0}$, and $n_\mathbf{p}$ is the total number of samples in subgroup $\mathbf{p}$.

\begin{center}
	\begin{tabular}{|c|c|c|c|c|}
		\hline
		& $Y=1$ & $Y=0$ & Total & \\
		\hline
		$W=1$ & $n_{11}$ & $n_{10}$ & $n_{1*}$ & $r_1 = n_{11} / n_{1*}$ \\
		$W=0$ & $n_{01}$ & $n_{00}$ & $n_{0*}$ & $r_0 = n_{01} / n_{0*}$ \\
		total & $n_{*1}$ & $n_{*0}$ & $n_{\mathbf{p}}$ & $\overline{r} = n_{*1} / n_{\mathbf{p}}$ \\
		\hline
	\end{tabular}
\end{center}

$\Delta= \lvert r_1 - r_0 \rvert $ is significantly grater than 0 if $z > z_c$ where $z = \frac{\lvert  r_1-r_0 \lvert  - \frac{1}{2} (\frac{1}{n_{1*}} + \frac{1}{n_{0*}})}{\sqrt{\overline{r}(1-\overline{r})(\frac{1}{n_{1*}} + \frac{1}{n_{0*}})}} $ and $z_c$ is the critical value at a confidence level. For example, when the confidence level is 95\%, $z_c =1.96$.  




The most specific TEPs and their general TEPs form a lattice in space $\mathbf{X}$. The number of TEPs can be large. We aim at finding the minimal set of TEPs that explain every individual with the most specific TEP. A TEP \emph{covers} a record in a data set if the TEP is a subset of the record when unspecified values in the TEP are dropped.     

\begin{definition}[The minimal significant TEP set]
	A TEP set is significant and minimal with respect to a data set when 1) each TEP is significant except that the most general TEP may be insignificant; 2) all TEPs in the set cover all records in the data set; and 3) each TEP is the most specific for some records, i.e. it covers at least one record that is not covered by another more specific TEP in the set. 
\end{definition}


The minimum in the above definition means non-redundancy. A more general TEP is redundant if it does not cover any new records in addition to its more specific TEPs. A redundant TEP is excluded from the  minimal significant TEP set. 
Figure~\ref{fig_exampl_TEP_Set} shows the minimal significant TEP set. Note that a record may be covered by more than one TEP.  For example, some records are covered by both TEPs $\{(0, *), 0\}$ and $\{(0, *), *\}$ (for $\pa'(Y) = \{(X_1, X_2), X_3\}$). We consider $\{(0, *), 0\}$ (the more specific one among two) is the TEP covering records. TEP $\{(0, *), *\}$ is not redundant since it covers records covered by TEP $\{(0, 0), 0\}$ which is not in the minimal significant TEP set. Note that it is possible that there are not enough significant TEPs to cover all instances in a data set and those uncovered instances are explained by the most general TEP corresponding to $\ate(W, Y)$. This is caused by the data limitation, and using $\ate(W, Y)$ to estimate treatment effect is reasonable.

Finding the minimal significant TEP set is to solve a set cover problem, which is NP-hard~\cite{ComplexityCoveringAlg}.  We will propose a greedy algorithm to find the minimal TEP set.

\subsection{TEP discovery via pattern generalisation}

We start with the set of most specific TEPs, some or all of which are insignificant. The main reason for an insignificant TEP is that the subgroup of the TEP is small. We will merge the subgroup with other subgroups to make the TEP of the merged subgroup significant. 

\begin{definition}[TEP generalisation]
	Generalisation is a merge process where one or more specified values are replaced by unspecified values `$*$'s. A generalised TEP represents two or more (if there are more than one unspecified value) most specific TEPs. 
\end{definition}

An example of TEP generalisation is given in Figure~\ref{fig_exampl_TEP_Set}. Patterns $\{(0,0),1\}$ and $\{(0,1),1\}$ (for $\pa'(Y) = \{(X_1, X_2), X_3\}$) are generalised as pattern $\{(0,*),1\}$. Patterns $\{(0,0),0\}$, $\{(0,*),0\}$ and $\{(0,1),1\}$ are generalised as patterns $\{(0,*),*\}$.

There are two constraints in the generalisation. 

\begin{enumerate}
	\item The generalisation should involve as small heterogeneity as possible. A generalised TEP denotes the number of subgroups represented by a set of most specific TEPs with different treatment effects. The differences between the treatment effects should be as small as possible to make the resulted causal effect represent the treatment effects of all subgroups well. 
	\item The generalisation should keep the specificity as high as possible. An unspecified value means a loss of specificity. The higher the speciality, the treatment effect represented by a TEP closer to the individual treatment effect. The lower the speciality, the treatment effect represented by a TEP closer to the average causal effect in the population. For the purpose of personalised decision making,  we wish a TEP is as specific as possible, and hence the number of `$*$' values should be minimised. A bottom up approach as proposed in this paper has an advantage over other existing top-down partition approaches to produce specific patterns.    
\end{enumerate}

We use the following measure to quantify the heterogeneity. 

\begin{definition}[Diversity]
	Let a generalised TEP $\mathbf{p}$ represent $k$ most specific TEPs, $\mathbf{p_1, p_2, \ldots, p_k}$, and $\theta$ be the average treatment effect of $k$ the TEPs. The diversity of treatment effect of pattern $\mathbf{p}$ is $\dv(\mathbf{p}) = \frac{1}{k}\sum_1^k (\cate_{\mathbf{p_k}} - \theta) $. 
\end{definition}

In the merge process, we prefer a merger with the smallest diversity. 


The specificity loss is measured by the number of `$*$' in a generalised TEP. To minimise the loss, the TEPs to be merged should have the smallest edit distance (or the number of different values).  

The generalisation can be modelled as a multiple objective optimisation problem following the two constraints. We design a level-wise generalisation algorithm by using the $\epsilon$-constraint strategy for a Pareto optimal solution~\cite{MultiobjectiveOptimization}. In each step, we constrain the specificity loss to the smallest possible loss, and search for the generalisation to minimise heterogeneity. More specifically, the search strategy is as the following.   

\begin{enumerate}
	\item for each insignificant pattern, find its closest patterns with the smallest edit distance (to maximise the specificity). 
	\item In the set of closest patterns, choose a pattern to generalise resulting in the smallest diversity in the generalised pattern (to minimise the heterogeneity). 
\end{enumerate}


Let diversity $dv_0$ be the diversity of the most general TEP in the data set. We do not merge patterns resulting in a diversity larger than $dv_0$ since in this case, the average treatment effect represents the individuals in the generalised pattern better. 

\subsection{Algorithm}
\label{sec_alg}

Based on the above discussions, we propose a DEEP algorithm to find the minimal set of significant TEPs in Algorithm~\ref{alg-deep}. The algorithm consists of three phases: Finding the local causal structure \{$\mathbf{Z, F}$\}; Initialisation of the most specific TEPs; and Generalising for discovering significant TEPs. After discussing the three phases, we will discuss the complexity of the algorithm and how to use TEPs for personalised decision making. 

\subsubsection{Finding the local causal structure \{$\mathbf{Z, F}$\} (Lines 1 - 7)}
Ideally, a causal DAG is given by domain experts, and $\mathbf{Z}$ and $\mathbf{F}$ are read from the DAG. However, in most applications, a causal DAG is unavailable.

For finding $\pa(Y)$ from data, one straightforward way is to learn an entire causal DAG from data. However, learning an entire DAG is computationally expensive or intractable with high dimensional data.  


Local structure discovery~\cite{Aliferis2010a}, i.e. discovering $\pc(Y)$, the set of Parents (direct causes) and Children (direct effects) of the target $Y$ is sufficient for our algorithm. In our problem setting, $Y$ does not have descendants, and hence, $\pc(Y)=\pa(Y)$. 
Several algorithms have been developed for discovering $\pc(Y)$, such as PC-Select~\cite{PC-select-2010}, MMPC (Max-Min Parents and Children)~\cite{Tsamardinos2006_MMPC} and  HITON-PC~\cite{Aliferis2003_Hiton}. 
These algorithms use the framework of constraint-based Bayesian network learning and employ conditional independence tests for finding the PC set of a variable. Their performance is very similar. We chose MMPC because of its newly updated implementation~\cite{MxM2017}. 
This is implemented in Line 1. 

To distinguish sets $\mathbf{Z}$ and $\mathbf{F}$ in $\pa'(Y)$, we use the following property to find $F \in \mathbf{F}$. $F \in \pa'(Y)$ is a parent of $Y$ only if $F  \indep W$ in data. This is because edges $(F, Y)$ and $(W, Y)$ form a collider at $Y$. This is implemented in Lines 2-7.

\subsubsection{Initialisation of the most specific TEPs (Lines 8 - 15)}
Three sets $\mathbf{S}, \overline{\mathbf{S}}$ and $\mathbf{A}$ initialised in Line 8 are used to store significant, insignificant and all TEPs, respectively. The data set is projected to variable set $\pa'(Y)$ in Line 9 since TEPs are defined in $\pa'(Y)$. Stratification is used to count the cross table for each pattern in Lines 10-11. 
The CATE of each TEP is calculated by its cross table in Line 12.  The significant patterns passing the statistical test are added to the TEP set in Line 13. The diversity of the most general  TEP is calculated in line 15 and assigned to $dv_0$.

\begin{algorithm}[th]
	\scriptsize
	\caption{Discovering trEatment Effect Patterns (DEEP)}
	\label{alg-deep}
	{\bf{Input}}: Data set $D$ of treatment variable $W$, outcome variable $Y$, and pretreatment variables $\mathbf{X}$; Confidence level for independence tests and significance test. \\
	{\bf{Output}}: the minimal TEP set $\mathbf{S}$.
	
	
	\begin{algorithmic}[1]
			\State call a local structure learning algorithm to find $PA(Y)$ 
			\State let $\mathbf{Z} = PA(Y) \setminus W$ and $\mathbf{F} = \emptyset$ 
			\For {each $X \in PA(Y)$}
			\If{$(X \indep W)$} 
			\State $\mathbf{Z}\leftarrow \mathbf{Z}\setminus \{X\}$ and $\mathbf{F} = \mathbf{F} \cup X$
			\EndIf
			\EndFor 
			\State let $\mathbf{S} = \emptyset$, $\overline{\mathbf{S}} = \emptyset$ and $\mathbf{A} = \emptyset$
			\State	project $D$ on $(W, \pa'(Y), Y)$ and save as $D'$
			\State stratify $D'$ by $\pa'(Y)$ and obtain all most specific patterns
			\State update cross tables for all most specific patterns
			\State calculate $\cate$ of all most specific patterns and add them to $\mathbf{A}$
			\State add significant patterns to $\mathbf{S}$ 
			\State let $\overline{\mathbf{S}} = \mathbf{A} - \mathbf{S}$ 
			\State let $dv_0$ be the diversity of the most general TEP
			\State calculate pairwise pattern distances between patterns in $\mathbf{A}$ and store the distances in matrix $\mathbf{M}$ 		
			\State let $d$ be the smallest distance in $\mathbf{M}$  		
			\While {($\overline{\mathbf{S}} \ne \emptyset$ and $d < \lvert \pa'(Y) \rvert$)}
			\State let $\mathbf{L}$ contain pattern pairs in $\mathbf{A}$ with pattern distance of $d$ and at least one pattern in each pair being insignificant
			\State generalise a pair with the closest CATEs to $\mathbf{p}$     
			\If {$DV_{\mathbf{p}} > dv_0$} 
			\State remove the pair from $\mathbf{L}$ and continue
			\EndIf	
			\State add pattern $\mathbf{p}$ to $\mathbf{A}$ and $\overline{\mathbf{S}}$ and remove the patterns used to generalise $\mathbf{p}$ from $\mathbf{A}$ and $\overline{\mathbf{S}}$ 
			\State remove all pairs in $\mathbf{L}$ used to generalising $\mathbf{p}$ 
			\If {pattern $\mathbf{p}$ is significant} 
			\State add $\mathbf{p}$ to $\mathbf{S}$ and remove $\mathbf{p}$ from $\overline{\mathbf{S}}$
			\EndIf 	
			\State update pairwise pattern distance matrix $\mathbf{M}$ 			  		
			\State let $d$ be the smallest value in $\mathbf{M}$  					
			\EndWhile
			\State add the most general pattern to $\mathbf{S}$ 
			\State output $\mathbf{S}$
	\end{algorithmic}
\end{algorithm}

\subsubsection{Generalising for discovering significant TEPs (Lines 16 - 32)}
Immediately after the above initialisation steps, all patterns in set $\mathbf{A}$ are most specific without the unspecified values `$*$'. Pairwise edit distances of all patterns are calculated and stored in matrix $\mathbf{M}$ in Line 16 and the shortest distance is found in Line 17. Note, in distance calculation, an unspecified value `$*$' and a specified value (1 or 0) are different. Two unspecified values are also different since they may represent different values. Lines 18-31 are for generalisation, and this process stops when $\overline{\mathbf{S}}$ is empty or TEPs in $\overline{\mathbf{S}}$ are nearly generalised to their most general form (only one specific value left). To prepare for generalisation, all pattern pairs with the shortest edit distance are found  and added to list $\mathbf{L}$ and the pairs involving both significant TEPs are excluded from the list since we aim at finding the minimal significant TEP set. In list $\mathbf{L}$, the pattern pair with the smallest difference among their treatment effects is generalised. If the diversity of the generalised pattern is larger than that of the most general TEP, $dv_0$. The generalised pattern is discarded and the pair is removed from $\mathbf{L}$. This is implemented in Lines 21-22. The TEPs used in the generalisation are replaced by the generalised pattern in both sets $\mathbf{A}$ and $\mathbf{\overline{S}}$. If the generalised pattern is significant, it is removed from the insignificant pattern set and added to the significant TEP set. The generalisation distance matrix is updated using the generalised patterns and the shortest pattern distance is found.   



After the loop, the most general pattern with all `$*$' values is added for those  uncovered records by TEPs in the data or coming test records that do not occur in the training data set. 


\subsubsection{Using TEPs for personalised decisions}
Significant TEPs identified from data are used for personalised decision making. Match an individual's record to the most specific TEP in the minimal significant TEP set. If more than one TEP match the record with the same specificity, the one with the largest $n$ (the cardinality of its covering set) is chosen. The treatment effect of the individual is estimated as the CATE of the TEP. The treatment is recommended to the individual if the CATE is positive, and the treatment is not recommended otherwise.   

\subsubsection{Time complexity}
Finding $\pa(Y)$ takes $O(\lvert \mathbf{X}\rvert \lvert \pa(Y)\rvert^{k+1})$ where $k$ is the size of the maximal conditional set for conditional independence test (usually $k=3-6$) by MMPC~\cite{Tsamardinos2006_MMPC,MxM2017}.  The initialisation of the most specific patterns takes $O(n\log(n))$  of time due to stratification. 
The pattern generalisation in the worst case takes $O(4^{\lvert \pa'(Y) \rvert})$ when all the most specific patterns are generalised, and in most cases, it takes less time. The overall time complexity is  $O(\lvert \mathbf{X}\rvert \lvert \pa(Y)^{k+1} \rvert + n\log(n) + O(4^{\vert \pa'(Y)\vert }))$. So the complexity is determined by the number of parents of the outcome variable. DEEP works for the data sets where the number of parents of the outcome is not many.   

\section{Experiments}\label{section_experiments}


\begin{table*}[t]
	\centering
	\caption{PEHE of different methods on 10 synthetics data sets using 10-cross validation. The standard deviation is shown in the parentheses. The smallest PEHE for each data set is highlighted.  }
	\footnotesize
	\begin{tabular}{l c c c c }		
		\hline
		\#Var &  CT & DEEP &  IT & UpliftDT  \\
		\hline
		20 &   0.201 (0.007) & \textbf{0.116} (0.005) & 0.154 (0.004) & 0.401 (0.070) \\
		40 &   0.240 (0.005) & \textbf{0.109} (0.005) & 0.171 (0.007) & 0.394 (0.008) \\
		60 &   0.257 (0.006) & \textbf{0.108} (0.003) & 0.214 (0.010) & 0.410 (0.007) \\
		80 &   0.278 (0.005) & \textbf{0.105} (0.003) & 0.267 (0.012) & 0.396(0.005) \\
		100 &   0.285 (0.007) & \textbf{0.107} (0.004) & 0.312 (0.014) & 0.400 (0.006) \\
		Ave &   0.252 (0.006) & \textbf{0.109} (0.004) & 0.223 (0.009) & 0.400 (0.019) \\
		\hline
	\end{tabular}
	\label{tab_PEHE}
\end{table*}

\begin{table}[t]
	\centering
	\footnotesize
	\caption{MAPE of different methods on 10 synthetics data sets using 10-cross validation. The standard deviation is shown in the parentheses. The smallest MAPE for each data set is highlighted.  }
	\begin{tabular}{l c c c c }
		\hline
		\#Var &  CT & DEEP &  IT & UpliftDT  \\
		\hline
		20 &   86.0 (4.3) & \textbf{51.4} (4.5) & 72.3 (3.5) & 246.1 (6.2) \\
		40 &   103.0 (2.8) & \textbf{45.0} (3.2) & 73.3 (4.3) & 212.8 (7.2) \\
		60 &   114.1 (3.7) & \textbf{47.2} (1.3) & 87.5 (6.5) & 234.9 (4.4) \\
		80 &   125.1 (4.3) & \textbf{46.4} (1.7) & 110.7 (6.3) & 255.2 (7.4) \\
		100 &   128.3 (5.1) & \textbf{45.4} (1.4) & 124.3 (9.9) & 224.5 (5.7) \\
		Ave &   111.3 (4.0) & \textbf{47.1} (2.6) & 93.6 (6.1) & 234.7 (6.2) \\
		\hline
	\end{tabular}
	\label{tab_MAPE}
\end{table}

\subsection{Baseline methods and parameter setting}
We compare DEEP with two state-of-the-art methods for causal effect heterogeneity modelling, Causal Tree (CT) \cite{AtheyImbens2016_PNAS}, and Interaction Tree (IT) \cite{su_subgroup_2009}, and one uplift modelling method, Uplift Decision Tree (UpliftDT) \cite{Rzepakowski2010_DTuplift}. 
All three methods are tree based, and their interpretability is comparable to DEEP's since a tree path can be interpreted as a pattern. Other causal heterogeneity and uplift modelling models do not provide the same interpretability and hence are not compared. 



We use the  CT implementation available at \url{https://github.com/susanathey/causalTree} by the authors of \cite{AtheyImbens2016_PNAS}. For IT, we use the R implementation available at \url{http://biopharmnet.com/subgroup-analysis-software/}. The default parameters are used for the two methods. UpliftTree is obtained from \url{https://causalml.readthedocs.io/en/latest/methodology.html#uplift-tree}. Euclidean distance is used since it performs best in the authors' work~\cite{Rzepakowski2010_DTuplift}. Other parameters are kept as the default. 

The parameters of DEEP are listed as follows. The confidence level for testing significant patterns in DEEP is set as 95\%. We have employed the R implementation of MMPC~\cite{MxM2017} for PC discovery, and set max$_k $ as 3, $p$ value as 0.05 and  gSquare for independence tests. 

\subsection{Evaluation of synthetic data sets}
This part aims at evaluating the quality of TEPs for modelling causal heterogeneity. The ground truth CATEs are necessary and hence the evaluation has been conducted in synthetic data sets. 

We have used the code in~\cite{AtheyImbens2016_PNAS} to generate synthetic data sets. Variables are binarised using their means since DEEP deals with binary variables. The numbers of variables are set as 20, 40, 60, 80 and 100 respectively, and the data set size is fixed at 10,000 for all. The number of parents of $Y$, i.e. $\lvert \mathbf{\mathbf{Z \cup F}} \rvert$, is 8 in all data sets. 10 data sets are generated randomly  in each setting. 

The ground truth CATEs are known and hence PEHE and MAPE are used for evaluating the quality of models. The Precision in Estimation of Heterogeneous Effects (PEHE) \cite{Hill2011} measures the mean squared error of estimated CATEs. i.e. $\textnormal{PEHE} = \frac{1}{n} \sum\limits_i^{n}(\hat{\tau}(\mathbf{x}_i)-\tau(\mathbf{x}_i))^2$ where $\hat{\tau}(\mathbf{x}_i)$ and $\tau(\mathbf{x}_i)$ are estimated CATE and ground truth CATE  of individual $\mathbf{x}_i$ respectively. The Mean Absolute Percentage Error (MAPE) is $\frac{1}{n} \sum\limits_{i}^{n} \vert \frac{\hat{\tau}(\mathbf{x}_i) - \tau(\mathbf{x}_i)}{\tau(\mathbf{x}_i)} \vert \times 100\%$. PEHE and MAPE are obtained by 10-cross validation in each data set and averaging over 10 data sets.

DEEP performs better than three other methods in terms of both PEHE and MAPE as shown in Tables~\ref{tab_PEHE} and~\ref{tab_MAPE}. This is because that DEEP keeps the information as specific as possible and hence predicts CATEs better than others.

\subsection{Evaluation on real world data sets}

We evaluate the methods on four real world data sets which are briefly described in Table~\ref{tab_datasets}. Criteo uplift prediction dataset~\cite{Diemert2018} is an open-access large scale data set. We have randomly sampled 200,000 records from the original data set. The Hillstrom's Email dataset is from~\url{https://blog.minethatdata.com/2008/03/minethatdata-e-mail-analytics-and-data.html}. The Marketing campaign data set is part of the Information R-package (\url{https://cran.r-project.org/web/packages/Information/index.html}). In the data sets, numerical variables have been binarised by their medians. The US Census (KDD) data set is from the UCI Machine Learning Repository~\cite{bache_uci_nodate}. We have selected the following attributes for easy interpretation: `College degree' (the treatment), `Income $>$ 50K' (the outcome), `Age $<$ 30', `Age $>$ 60', `Work-in-Private", `Work-in-Government", `Self-employed', `Professional', and `Full time', and `Sex'.

\begin{table*}[t]
	\centering
	\caption{A brief description of the real world data sets}
	\label{tab_datasets}
\footnotesize
	\begin{tabular}{ccccc}
		\hline
	Name	 & Criteo    & Hillstrom's     & Marketing    & US   \\
		 &    & Email    & Campaign    & Census  \\
		\hline
		
	\#Records	    &    200000    &   42693    &    20000     &    348128  \\
	\#Var	    &    12    &    29    &    67    &    8 \\
	Treatment	& Promotional    &  Women's      &    Marketing    &    College     \\
		& email    &  email     &    offer    &   degree     \\
	Outcome	 &  Visit    &  Visit      &   Accept     &       Income$\ge$50K  \\
	 &  (4.7\%)   &  (12.9\%)     &   (20.0\%)    &       (8.7\%) \\		
		\hline
	\end{tabular}
\end{table*}

\begin{figure*}[!t]
	\center
	\footnotesize
	\begin{tabular}{cccc}
		Criteo  & Hillstrom's Email	& Market Campaign & US Census\\
		\raisebox{-.5\height}{\includegraphics[width=0.23 \textwidth]{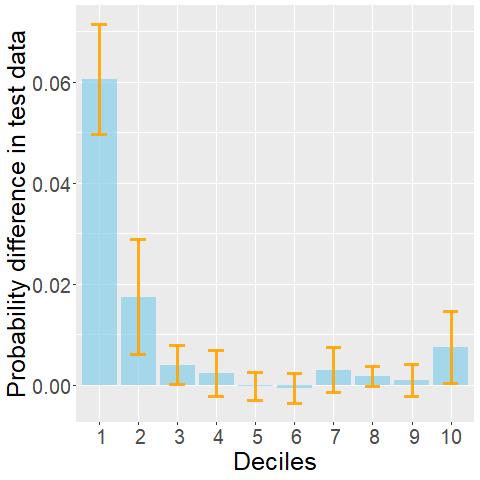}} & 	\raisebox{-.5\height}{\includegraphics[width=0.23 \textwidth]{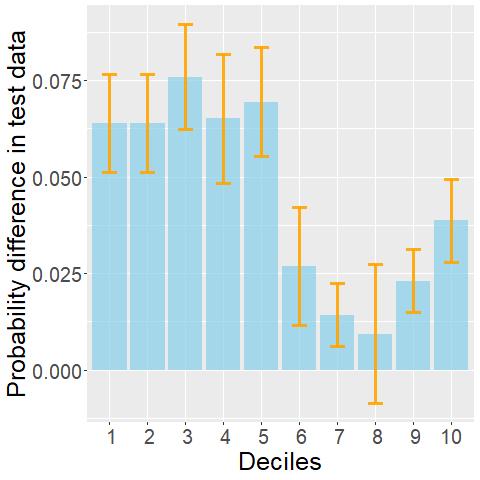}} & \raisebox{-.5\height}{\includegraphics[width=0.23 \textwidth]{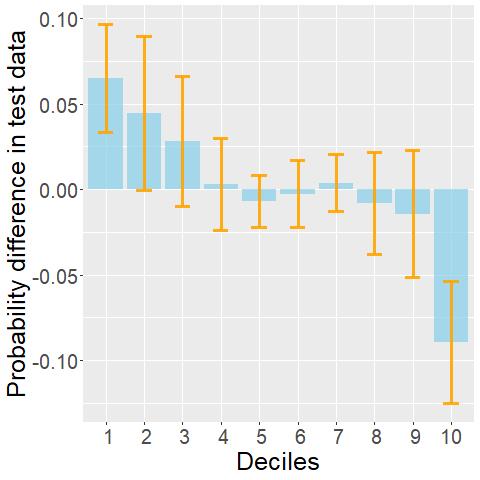}} & \raisebox{-.5\height}{\includegraphics[width=0.23 \textwidth]{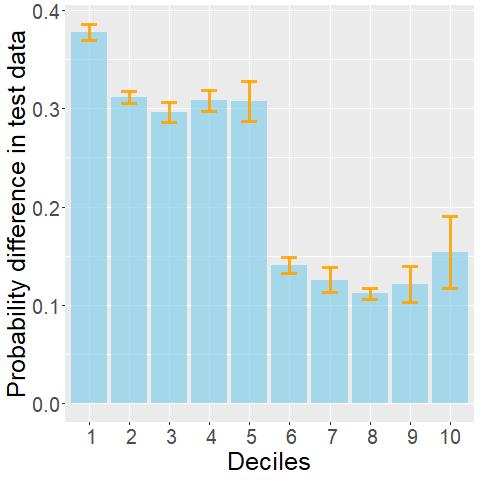}} \\
		\raisebox{-.5\height}{\includegraphics[width=0.23 \textwidth]{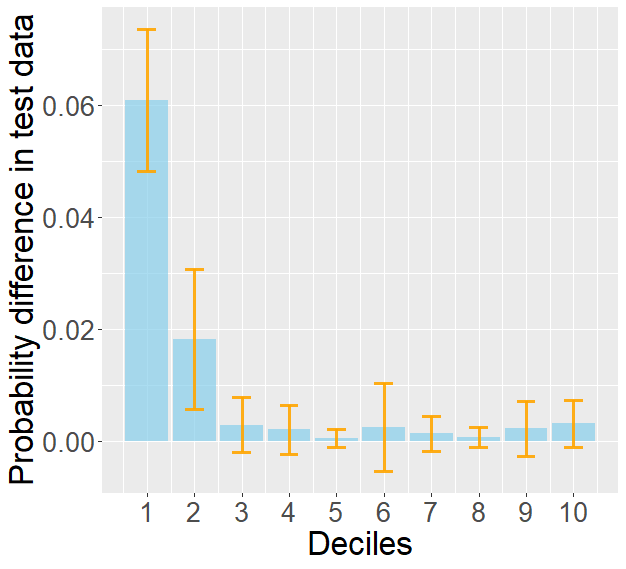}} &   	 \raisebox{-.5\height}{\includegraphics[width=0.23 \textwidth]{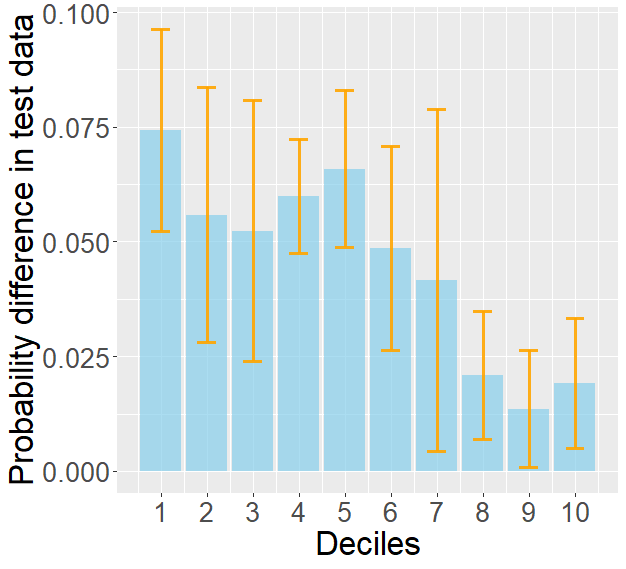}} &   	\raisebox{-.5\height}{\includegraphics[width=0.23 \textwidth]{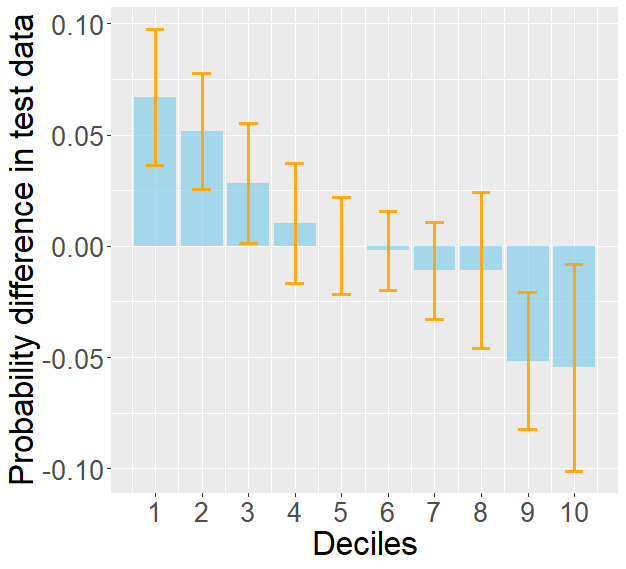}} & \raisebox{-.5\height}{\includegraphics[width=0.23 \textwidth]{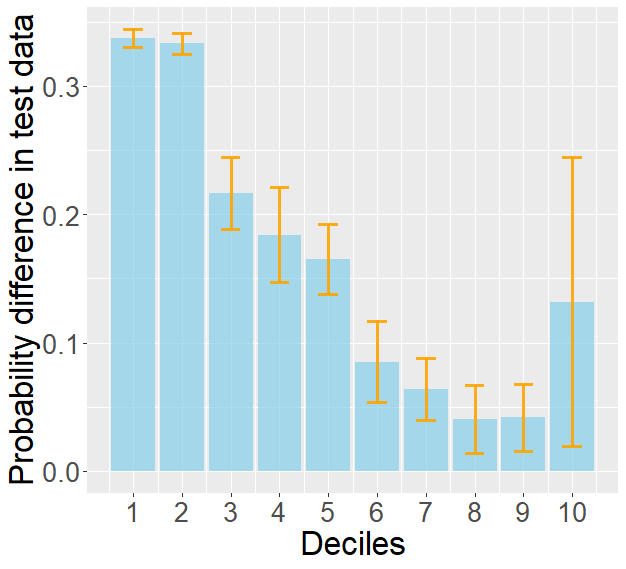}} \\
		\raisebox{-.5\height}{\includegraphics[width=0.23 \textwidth]{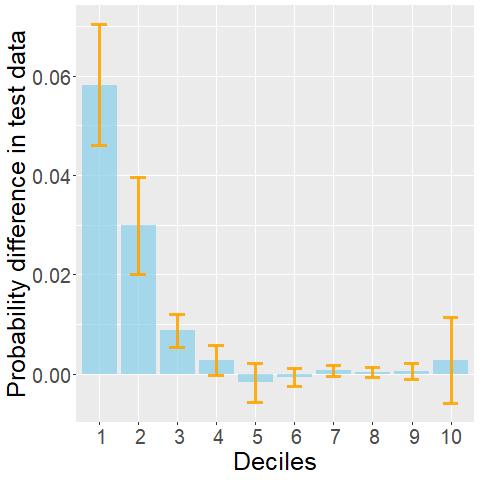}} &   	  \raisebox{-.5\height}{\includegraphics[width=0.23 \textwidth]{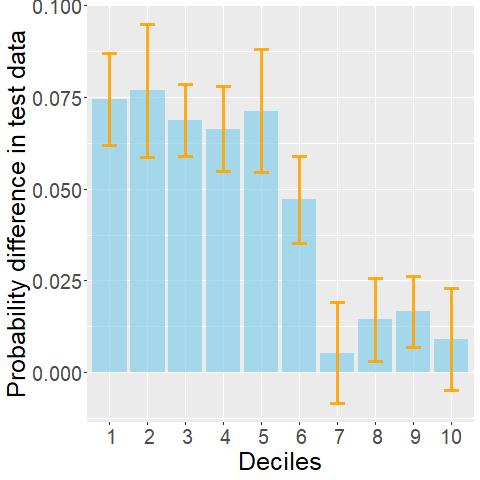}} &   	  \raisebox{-.5\height}{\includegraphics[width=0.23 \textwidth]{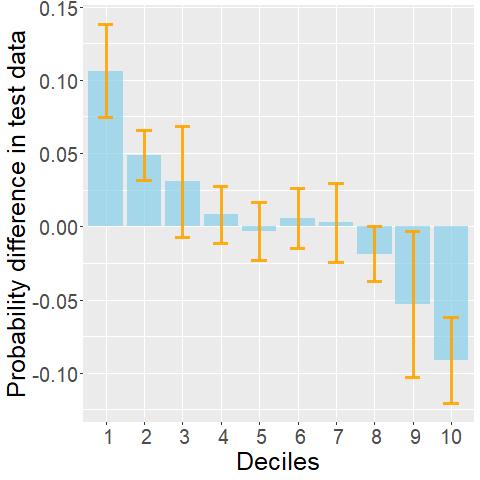}} & \raisebox{-.5\height}{\includegraphics[width=0.23 \textwidth]{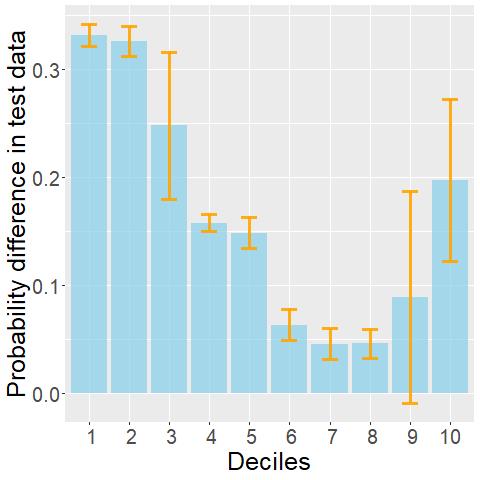}} \\ 	 \raisebox{-.5\height}{\includegraphics[width=0.23 \textwidth]{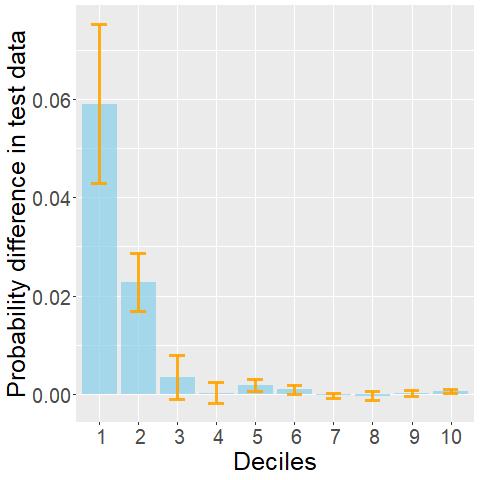}} &   	\raisebox{-.5\height}{\includegraphics[width=0.23 \textwidth]{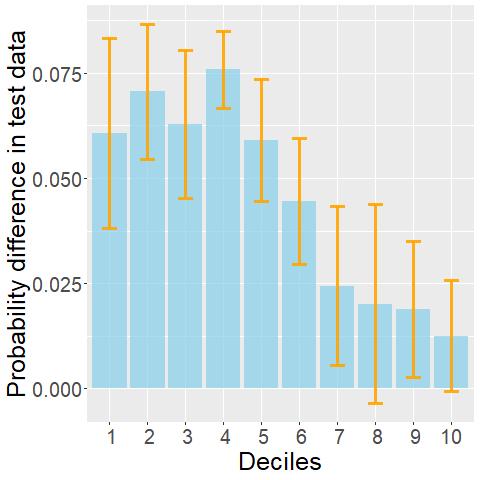}} &   	\raisebox{-.5\height}{\includegraphics[width=0.23 \textwidth]{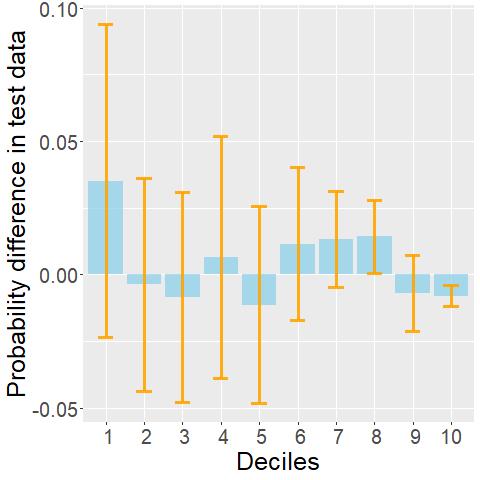}} & \raisebox{-.5\height}{\includegraphics[width=0.23 \textwidth]{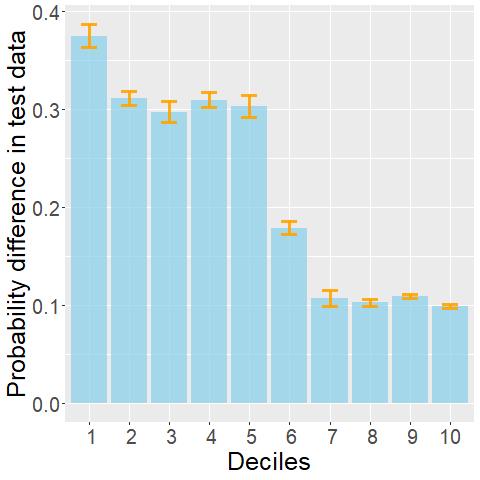}} \\
	\end{tabular}
	\caption{Decile plots of four methods in four data sets. From the 1st row to the last: CT, DEEP, IT and UplifDT. DEEP shows a consistent decline in four decile plots. } 
	\label{fig_Decileplot}
\end{figure*}

Since there are no ground truth CATEs in the real world data sets, we cannot use PEHE and MAPE to assess the quality of the methods. Instead, we use prediction accuracy for the assessment. A predicted CATE indicates the chance of improvement of an individual if s/he takes the treatment. We cannot assess the accuracy of each prediction. However, we can estimate the cumulative improvement of a group of individuals. In a test data set, all individuals are ranked by their predicted CATEs, and are then partitioned into 10 groups: Decile 1 to 10 groups with CATE in the descending order. If a model is good, the observed difference, $\prob(y \mid W=1) - \prob(y \mid W=0)$, in the 10 groups will monotonically decrease with the increase of the decile indexes. The higher quality a model is, the steeper the declining rate. The decile plots have been used for assessing the quality of uplift models~\cite{Gutierrez2017_CausalUpliftReview}. 

The decile plots in Figure~\ref{fig_Decileplot} show that DEEP performs overall more consistent than other methods. In data sets Criteo, Hillstrom's email and US Census, DEEP performs better than others since it presents a steep declining curve. In the Market Campaign data set, DEEP's performance is very competitive with CT and IT and better than UpliftDT. No other algorithms perform as consistent as DEEP in all four data sets. The results have been obtained by 10 times 2-fold cross validation.

\begin{table*}[t]
	\centering
	\caption{The number of patterns (or paths from the root to leaves in a tree) from different methods in four data sets. }
\footnotesize
	\begin{tabular}{l c c c c }
		\hline
		&  CT & DEEP &  IT & UpliftDT  \\
		\hline
		Criteo & 14.3 (4.6) & 39.4 (6.1) & 4.7 (1.3) & 31.8 (2.5) \\
		Hillstrom & 47.5 (7.9) & 24.9 (3.8) & 2.4 (1.0) & 38.7 (4.6) \\
		Market &   75.1  (11.0) & 58.6 (13.5) & 10.4 (4.1) & 30.5 (2.3) \\
		US Census & 8.7 (2.4) & 75.0 (1.7) & 24.3 (2.1) & 3.7 (1.9) \\
		\hline
	\end{tabular}
	\label{tab_patternNum}
\end{table*}

The number of patterns (or paths from the root to leaves) are shown in Table~\ref{tab_patternNum}. DEEP does not discover too many patterns and this is due to the significant test for a pattern. A tree based method is able to find many subgroups by increasing the tree height, but a tree based method does not have flexibility like DEEP since all patterns from a tree are constrained by the variable at the root: all patterns include a value of the root variable. In contrast, DEEP does not have such a constraint and can model any heterogeneous subgroups.


\begin{figure*}[t!]
	\centering
	\begin{minipage}{0.4\textwidth}
		\centering
		\includegraphics[width=\textwidth]{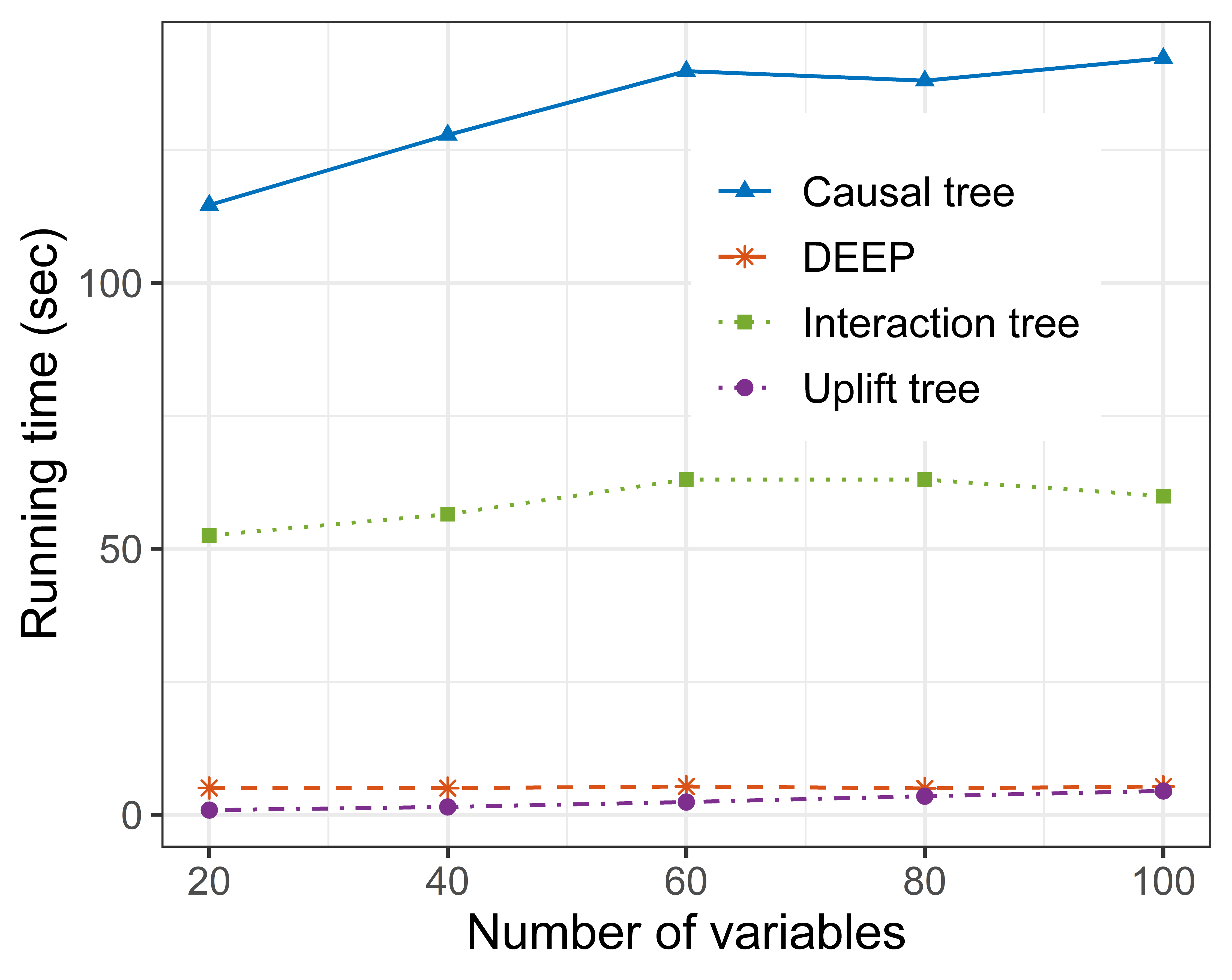}
	\end{minipage}
	\begin{minipage}{0.4\textwidth}
		\centering
		\includegraphics[width=\textwidth]{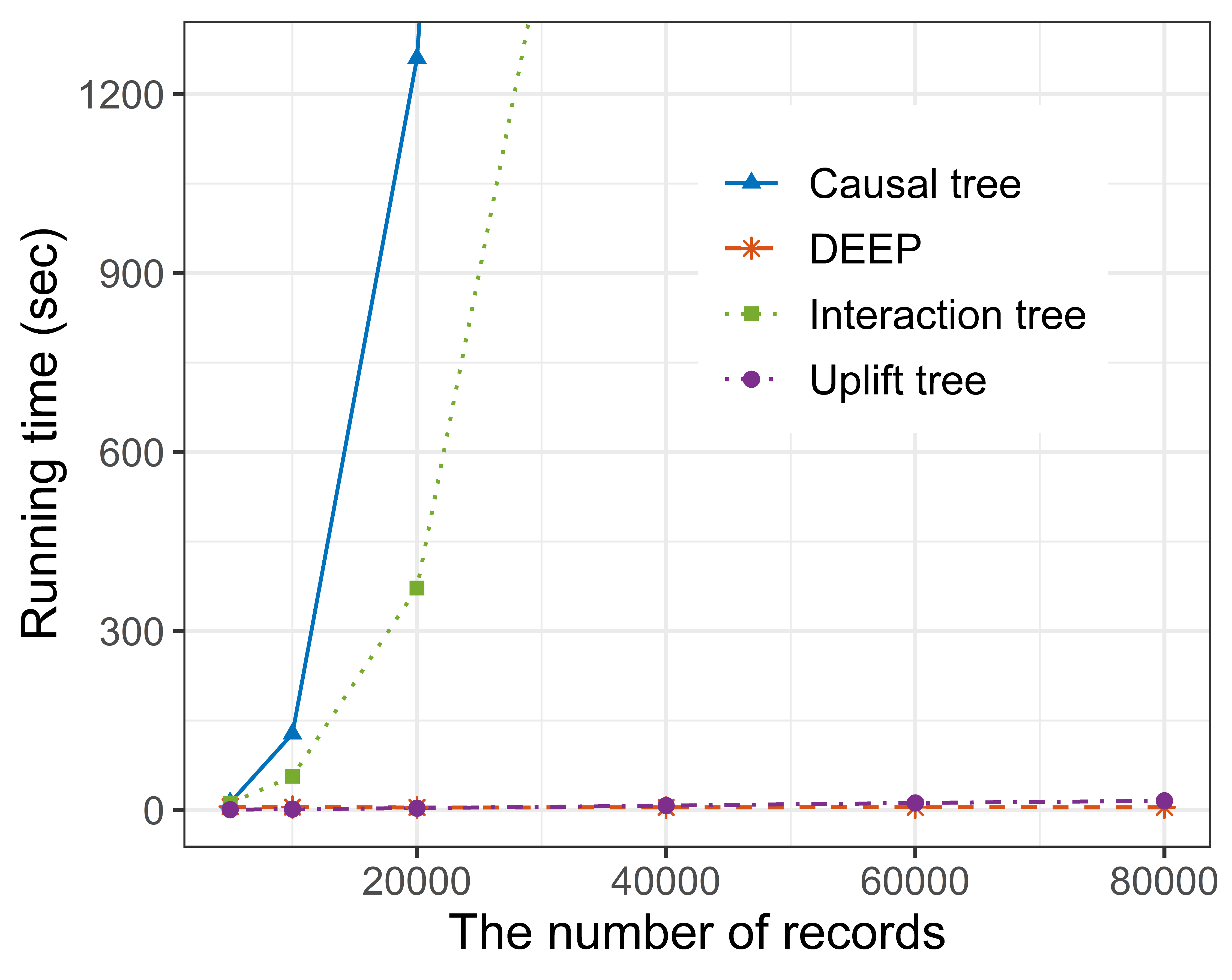}
	\end{minipage}
	\caption{Time efficiency with the number of variables (left), and with the number of records (right) of four methods.}
	\label{fig_efficiency}
\end{figure*}

\subsection{Time efficiency}
We apply DEEP and three other methods to the synthetic data sets of 20, 40, 60, 80, and 100 variables with 10,000 records for testing their scalability with the number of variables, and to the synthetic data sets of 5K, 10K, 20K, 40K, 60K, and 80K with 40 variables for testing their scalability with the data set size. Results are shown in Figure~\ref{fig_efficiency}. 

With the number of variables, the scalability of all four methods is good. Relatively, UpliftDT is the fastest since it does not estimate CATEs or test reliability for tree splits. CT is the slowest since it needs to estimate propensity scores for CATE estimation. Logics regression is used in the propensity score estimation. DEEP and IT perform similarly.

With the number of records, DEEP and UpliftDT perform very well. The increase in data set size improves the time efficiency of DEEP since pattern generalisation is an expensive part of DEEP. With the increase in data set size, the number of significant patterns at the most specific level is increasing, and this reduces the overall number of patterns to be merged. Since CT uses logistic regression for propensity score estimation, its performance deteriorates quickly with the increase in data set size. IT grows a large tree firstly and then prunes it back to a small tree. In the pruning process, cross validation is used to determine whether to retain a leaf node or not, and this leads to its low scalability with the data set size.  The scalability of DEEP the data set size is good. DEEP works for large sized data sets.

\section{Related work}\label{section_related_work}

Great research efforts have been made on treatment effect estimation within two major frameworks: graphic causal modelling~\cite{Pearl2009_Book} and potential outcome modelling~\cite{ImbensRubin2015_Book}. The work in this paper falls into the former. 


CATEs are commonly analysed to detect treatment effect heterogeneity and we are interested in data driven analysis. Su et al. \cite{su_subgroup_2009} used recursive partitioning to construct the interaction tree (IT) for treatment effect estimation in different subgroups by adapting the CART~\cite{CART1984}. Athey  et al. \cite{AtheyImbens2016_PNAS} proposed to use honest estimation for tree partition and causal effect estimation, and built the Causal Tree (CT) based on the CART~\cite{CART1984} to find the subpopulations with heterogeneous treatment effects. Wager and Athey further proposed a random forest based method for causal effect heterogeneity modelling~\cite{WagerAthey2018_RF}. A meta-learning method~\cite{kunzel2019metalearners} was proposed for causal heterogeneity modelling with unbalanced treated and control samples. In recent years, some algorithms have been presented using deep learning techniques~\cite{Shalit2016,yoon2018ganite,CEVAE2017,Yao2018_Twin}. Interesting readers are referred to a survey~\cite{CausalSurvey2020} and an evaluation paper~\cite{CHM-Evaluation}.



Uplift modelling is closely linked to causal heterogeneity modelling as shown in~\cite{Gutierrez2017_CausalUpliftReview,zhang2020unified}. Due to the page limit, we refer readers to the recent surveys~\cite{DevriendtSurvey2018,GubelaEvaluation2019}. Uplift modelling is normally assumed in data from a well designed randomised experiment and hence probability difference in the treated and control groups has been used as CART without adjustment. Therefore, it is not clear whether the uplift modelling methods can be used in observational data. Again only tree based methods are of our interest because of the interpretability. Rzepakowski and Jaroszewicz adapted decision trees for uplift modelling~\cite{Rzepakowski2010_DTuplift,Rzepakowski2012_DTforUpliftModel}.

A covariate set in causal inference should satisfy the unconfoundedness assumption (i.e. conditional ignorability~\cite{rosenbaum1983central}). 
VanderWeele and Shpitser~\cite{VanderWeele-NewCriteria} have proposed a covariate set to be the union of causes of the treatment and causes of the outcome without knowing the underlying causal structure. de Luna et al.~\cite{de2011covariate} have proposed a method to reduce a covariate set to the minimal sets under the unconfoundedness assumption, and an implementation based on the Bayesian network has been reported in~\cite{haggstrom2018data}.  Entner et al.~\cite{entner2013data} have proposed a method to find covariate sets using conditional independence tests. 
These works focus on ATE estimation instead of CATE estimation and they have not elaborated on the role of confounders and effect modifiers in CATE estimation. PC (parent and child) discovery algorithms, such as PC-Select~\cite{PC-select-2010}, MMPC (Max-Min Parents and Children)~\cite{Tsamardinos2006_MMPC} and HITON-PC~\cite{Aliferis2003_Hiton}, can be considered covariate selection algorithms when data sets contain pretreatment variables (both ancestral nodes of treatment and the outcome in a causal graph term).

Causal rules~\cite{li2015observational,Jin2012_ICDM} and causal patterns~\cite{yadav2019frequent} concern multiple treatments, not causal heterogeneity. They are not relevant 


\section{Conclusions}\label{section_conclusion}
We have proposed TEPs to represent treatment effect heterogeneity in a population. TEPs encode the local causal structure which gives users an overview of causal relationships around the outcome variable. Users can evaluate TEPs discovered in data  based on the consistency between the local causal structure and their domain knowledge, and can also use their believed local causal structure to guide TEP discovery. We have developed the DEEP algorithm to identify TEPs using a bottom up approach which ensures that each TEP is as specific as possible while its subgroup has the smallest possible treatment effect heterogeneity. When using the discovered TEPs, the most specific TEP matching a person's situation is used for personalised decision making. The experiments show that the DEEP models the treatment heterogeneity better than three existing tree based methods in both synthetic and real world data sets and DEEP is efficient among the comparison methods.  
Our future work will apply the DEEP to assist personalised decision making in various applications and extend the TEP for other types of variables other than binary variables. 

\section*{Acknowledgement}
	This work has been supported by the Australian Research Council [grant number: DP200101210 and DE200100200].
	


\bibliographystyle{acm}
\bibliography{CausalHeterogenerityJan2021}

	\end{document}